\documentclass[aps,prl,twocolumn,showpacs,amsmath,amssymb,floatfix]{revtex4}
\usepackage{epsfig}
\usepackage{bm}

\begin{document}

\title{Resonant dephasing in the electronic Mach-Zehnder interferometer.}

\author{Eugene V. Sukhorukov$^1$, and Vadim V. Cheianov$^2$}
\affiliation{$^1$\
D\'{e}partment de Physique Th\'{e}orique, Universit\'{e} de Gen\`{e}ve,
CH-1211 Gen\`{e}ve 4, Switzerland\\
$^2$\ Physics Department, Lancaster University, Lancaster, LA1 4YB, UK
}
\date{\today}

\begin{abstract}
We address the recently-observed unexpected behavior of
Aharonov-Bohm oscillations in the electronic Mach-Zehnder
interferometer that was realized experimentally in a quantum Hall
system \cite{Heiblum1}. We argue that the measured lobe structure in
the visibility of oscillations and the phase rigidity result from  a
strong long-range interaction between two adjacent
counter-propagating edge states, which leads to a resonant
scattering of plasmons.   The visibility and phase shift, which we
expressed in terms of the transmission coefficient for plasmons, can
be used for the tomography of edge states.
\end{abstract}

\pacs{73.23.-b, 85.35.Ds, 03.65.Yz}

\maketitle

Quantum interference effects, particularly the Aharonov-Bohm (AB)
effect \cite{Aharonov1}, and their suppression due to interactions
\cite{Dephasing} have always been a central subject of mesoscopic
physics, and by now are thoroughly investigated. However, recent
experiments on the AB effect in Mach-Zehnder (MZ)
\cite{Heiblum1,Heiblum2} and Fabry-Perot type \cite{Goldman}
interferometers, which utilize quantum Hall edge states \cite{QHE}
as one-dimensional conductors, have posed a number of puzzles
indicating that the physics of edge states is not yet well
understood \cite{IV}. For instance the lobe-type pattern in the
visibility of AB oscillations as a function of voltage bias, as well
as the rigidity of the phase of oscillations followed by abrupt
jumps by $\pi$, observed in Ref.\ \cite{Heiblum1}, cannot be
explained within the single-particle formalism \cite{Buttiker} which
is supposed to describe edge states at integer filling factor
\cite{Wen}.

\begin{figure}[htb]
\epsfxsize=8cm
\epsfbox{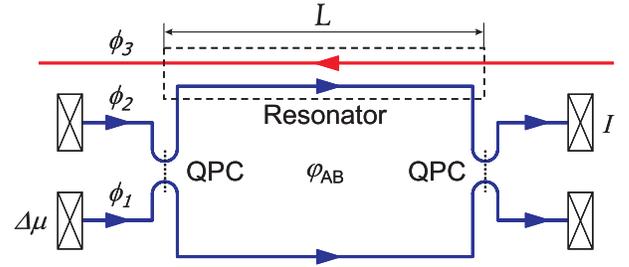} \caption{Schematic representation of the
electronic Mach-Zehnder (MZ) interferometer experimentally realized
in Ref.\ \cite{Heiblum1}. Two edge states (blue lines, $\phi_1$ and $\phi_2$), 
that propagate from left to right, are coupled via two quantum point
contacts (QPC) and form the  Aharonov-Bohm  loop. The bias
$\Delta\mu$ applied to one of the source ohmic contacts causes the
current $I$ to flow around the loop, so that the differential
conductance $G=dI/d\Delta\mu$ oscillates as a function of the phase
$\varphi_{\rm AB}$. The specific property of the set-up in Ref.\
\cite{Heiblum1} is that the third counter-propagating edge state
(red line, $\phi_3$) closely approaches the upper branch of the MZ
interferometer (inside the resonator shown by a dashed box) and
strongly interacts with it.} \vspace{-5mm} \label{setup}
\end{figure}

Indeed, according to a single-particle picture the electron edge
states propagate as plane waves with the group velocity $v_F$ at
Fermi level. They are transmitted through the MZ interferometer (see
Fig.\ \ref{setup}) at the left and right quantum point contacts
(QPC) with amplitudes $t_L$ and $t_R$, respectively. In the case of
low transmission, two amplitudes add so that the total transmission
probability $T\propto |t_L|^2+|t_R|^2+2|t_Lt_R|\cos(\varphi_{\rm
AB}+ \Delta\mu\Delta x/v_F)$ oscillates as a function of the AB
phase $\varphi_{\rm AB}$ and bias $\Delta\mu$, where $\Delta x$ is
the length difference between two interfering paths of the
interferometer. The AB oscillations may be seen in the differential
conductance $G=dI/d\Delta\mu$, which is given by the
Landauer-B\"uttiker formula \cite{Buttiker}, $G=T/2\pi$. The degree
of coherence is quantified by the visibility of AB oscillations
$V_{\rm AB}=(G_{\rm max}-G_{\rm min})/(G_{\rm max}+G_{\rm min})$,
which for low transmission acquires the simple form
\begin{equation}
V^{(0)}_{\rm AB}=2|t_Lt_R|/(|t_L|^2+|t_R|^2),
\label{noninteracting-VAB}
\end{equation}
i.e.\ it is independent of bias. Moreover,
the phase shift of AB oscillations is just
a linear function of bias, $\Delta\varphi_{\rm AB}=\Delta\mu\Delta x/v_F$.

Dephasing in ballistic mesoscopic rings was reported in Ref.\
\cite{Hansen} and theoretically addressed in Ref.\ \cite{Seelig}.
Since the first experiment on an electronic MZ interferometer
\cite{Heiblum2}, several theoretical models of dephasing in this
particular system have been proposed, including classical
fluctuating field \cite{Marquardt1} and dephasing probe \cite{Chung}
models. However, the unusual dephasing of AB oscillations in Ref.\
\cite{Heiblum1} (see also \cite{Florian}) seems to arise from a
specific interaction at the edge of a quantum Hall system.

\begin{figure}
\epsfxsize=7.5cm
\epsfbox{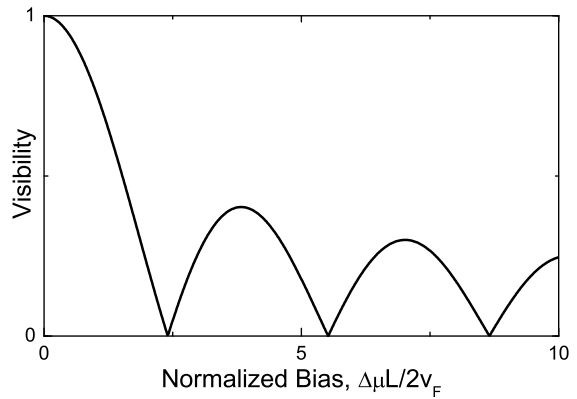}
\caption{In the case of a long-range interaction of counter-propagating
edge states at the resonator of the length $L$ (see Fig.\ \ref{setup}),
the visibility of Aharonov-Bohm (AB)
oscillations varies as a function of the normalized bias, $\Delta\mu L/2v_F$,
in a lobe-like manner. The phase of AB oscillations (not shown) stays
constant at the lobes and changes abruptly by $\pi$ at zeros of the
visibility. The visibility is plotted here for the simplified capacitive
coupling model, Eq.\ (\ref{Long-range}), with $T_L=T_R$.}
\vspace{-5mm}
\label{Bessel}
\end{figure}

In this Letter we propose a model which may explain the unusual AB
effect. We note that an important feature of the MZ setup
\cite{Heiblum1} is the existence of a counter-propagating edge state
(labeled as $\phi_3$ in Fig.\ \ref{setup}), which closely approaches
the edge state forming the upper arm of
the interferometer (labeled as $\phi_2$ in Fig.\ \ref{setup}) and
strongly interacts with it \cite{footnote}. Being localized inside a
finite interval of the length $L$, the interaction leads to a
resonant scattering of collective charge excitations (plasmons),
which carry away the phase information. As a result, AB oscillations
vanish at certain values of bias $\Delta\mu$ (see Fig.\
\ref{Bessel}), where the AB phase jumps by $\pi$. We found an
important relation between the transmission coefficient of plasmons
and the visibility of AB oscillation, equations
(\ref{Visibility-phase}) and (\ref{Result}), which opens a
possibility for the tomography of the edge state interactions.

{\it Model of a Mach-Zehnder interferometer.}--- To describe quantum
Hall edges at filling factor $\nu=1$, we apply the chiral Luttinger
liquid model \cite{Wen,Giamarchi} and write the Hamiltonian as
$H=(v_F/4\pi)\sum_\alpha\int dx\,[\nabla \phi_\alpha(x)]^2+H_{\rm
int}$. The bosonic fields $\phi_\alpha$, $\alpha=1,2,3$, describe
low-energy collective charge excitations at the edges,
$\rho_\alpha(x)=(1/2\pi)\nabla\phi_\alpha(x)$. They satisfy the
commutation relations: $[\phi_\alpha(x),\phi_\alpha(y)]=\pm i\pi{\rm
sgn}(x-y)$, where the minus sign stands for the counter-propagating
field $\phi_3$. Details of the interaction $H_{\rm int}$ are not
exactly known. Here we assume a general density-density interaction:
$H_{\rm int}=(1/2)\sum_{\alpha\beta}\iint dxdy\, U_{\alpha\beta}
(x,y)\rho_\alpha(x)\rho_\beta(y)$.

We note that electron-electron interaction within one edge, while
generally leading to a smooth suppression of the visibility
\cite{Chalker} as a function of $\Delta\mu$, cannot explain the lobe
structure observed in Ref.\ \cite{Heiblum1}. We therefore further
assume that only the elements $U_{23}$ and $U_{22}=U_{33}$ inside
the resonator are nonzero. Going over to the canonical variable
$\phi=(\phi_2+\phi_3)/2$ and its dual variable
$\theta=(\phi_3-\phi_2)/2$, such that
$[\phi(x),\nabla\theta(y)]=i\pi\delta(x-y)$, we obtain the final
Hamiltonian $H=(v/4\pi)\int dx\,[\nabla \phi_1(x)]^2+H_{\rm LL}$,
where the important part
\begin{align}
H_{\rm LL}&=
\frac{v_F}{2\pi}\int dx[(\nabla\phi)^2+(\nabla\theta)^2]\notag\\
& +\iint \frac{dxdy}{4\pi^2}
[U(x,y)\nabla\phi(x)\nabla\phi(y)\notag\\
&\qquad
\qquad
\qquad
\qquad
\quad
+V(x,y)\nabla\theta(x)\nabla\theta(y)],
\label{LL}
\end{align}
takes into account the interaction,
$U\equiv U_{22}+U_{23}$, $V\equiv U_{22}-U_{23}$, at the resonator.

In the experiment \cite{Heiblum1}, two point contacts located at
$x_\ell$, $\ell=L,R$, mix the edge states and allow interference
between them. This can be described by the tunneling Hamiltonian
\cite{footnote1}
\begin{subequations}
\begin{eqnarray}
H_T &=& A+A^{\dagger},\quad
A=A_L+A_R,
\label{Tunneling1}\\
A_\ell &=&t_\ell\psi^{\dagger}_2(x_\ell)\psi_1(x_\ell),\quad \ell=L,R,
\label{Tunneling2}
\end{eqnarray}
\label{Tunneling}
\end{subequations}
where $\psi_{1}$ and $\psi_{2}$ are electron operators, and the
tunneling amplitudes $t_\ell$ depend on the Aharonov-Bohm phase
$\varphi_{\rm AB}$.

{\it Visibility of Aharonov-Bohm oscillations.}---
We will investigate interference effects in the tunneling current
$I=i(A-A^\dagger)$, see Fig.\ \ref{setup}.
To the lowest order in tunneling amplitudes $t_{\ell}$
its expectation value is given by
$I=\int dt\, \langle[A^\dagger(t),A(0)]\rangle$ \cite{footnote2},
where the average is taken with respect to
the ground state of the system biased by the
potential difference $\Delta\mu$. Taking into account Eqs.\
(\ref{Tunneling}), we write the total current as a sum of three
terms: $I=I_L+I_R+I_{LR}$, where two terms
$I_\ell=\int dt\, \langle[A_\ell^\dagger(t),A_\ell(0)]\rangle$
are direct contributions of two point contacts, and
$I_{LR}=\int dt\, \langle[A_L^\dagger(t),A_R(0)]\rangle+c.\,c.$ is the
interference term that contains the AB phase.

Next, we recall that in our model the interaction is effectively
present only at the resonator between points $x_L$ and $x_R$. There
are two important consequences of this. First, the interaction
cannot affect direct contributions $I_L$ and $I_R$. Therefore, we
readily obtain the conductances for non-interacting electrons:
$dI_{\ell}/d\Delta\mu=G_\ell=2\pi n_F^2|t_\ell|^2$, where $n_F$ is
the density of states at the Fermi level. And second, the
interference term, $G_{LR}=n_Ft_Lt_R\exp(i\Delta\mu\Delta x/v_F)
{\cal G}_{\Delta\mu} +c.\,c.$, still depends on the interaction via
the Fourier transform ${\cal G}_{\Delta\mu}=v_F^{-1}\int dX
\exp[i(\Delta\mu/v_F)X]\,{\cal G}(X)$ of the electronic correlator
\begin{equation}
{\cal G}(X)=\langle\psi_2(x_L,t)\psi^{\dagger}_2(x_R)\rangle,
\label{DEF}
\end{equation}
which, however, depends on coordinates only via the combination $X\equiv x_R-x_L+v_Ft$.
Thus we obtain an important result for the visibility of AB oscillations, and for the
AB phase shift:
\begin{subequations}
\begin{eqnarray}
V_{\rm AB}/V^{(0)}_{\rm AB}=(1/2\pi n_F)\,|{\cal G}_{\Delta\mu}|,
\label{Visibility}\\
\Delta\varphi_{\rm AB}=\Delta\mu\Delta x/v_F+\arg\,({\cal G}_{\Delta\mu}),
\label{Phase}
\end{eqnarray}
\label{Visibility-phase}
\end{subequations}
where $V^{(0)}_{\rm AB}$ is the visibility
in the absence of interaction, see Eq.\ (\ref{noninteracting-VAB}).

{\it Electron correlation function.}---
As a next step, we quantize plasmons, taking into account inhomogeneous
interaction. The Hamiltonian (\ref{LL}) generates two coupled equations
of motion for fields $\phi$ and $\theta$.
We choose
periodic boundary conditions on the spatial interval of the length $W$,
which in the end is taken to infinity. Then the equations of motion may
be solved in terms of an infinite set $\{\Phi_n,\Theta_n\}$
of mutually orthogonal eigenfunctions which satisfy equations
\begin{subequations}
\begin{align}
\omega_n \Phi_n+v_F\nabla \Theta_n
&=-(2\pi)^{-1}\!\!\int dy\,V(x,y)\nabla \Theta_n(y),
\label{Motion1}\\
\omega_n \Theta_n-v_F\nabla \Phi_n
&=(2\pi)^{-1}\!\!\int dy\,U(x,y)\nabla \Phi_n(y),
\label{Motion2}
\end{align}
\label{Motion}
\end{subequations}
and can be chosen to be real and normalized as follows:
\begin{equation}
\int_0^W\! dx\,\Phi_n(x)\nabla \Theta_m(x)=-\omega_n\delta_{nm}.
\label{Normalization}
\end{equation}
The solutions then read
\begin{subequations}
\begin{align}
\phi(x,t)=&\sum_n\sqrt{\frac{\pi}{2\omega_n}}\,\Phi_n(x)
\left(a_ne^{-i\omega_nt}+a^{\dagger}_ne^{i\omega_nt}\right),
\label{Solution1}\\
\theta(x,t)=i&\sum_n\sqrt{\frac{\pi}{2\omega_n}}\,\Theta_n(x)
\left(a_ne^{-i\omega_nt}-a^{\dagger}_ne^{i\omega_nt}\right),
\label{Solution2}
\end{align}
\label{Solution}
\end{subequations}
where the plasmon operators $a_n$ satisfy
the commutation relations $[a_n,a^{\dagger}_m]=\delta_{nm}$ and diagonalize
the Hamiltonian: $H_{LL}=\sum_n\omega_n a_n^{\dagger}a_n$.

Proceeding with the bosonization of the electron operators, we write
$\psi_2\propto\exp[i(\phi-\theta)]$, where the normalization
prefactor is determined by the ultra-violet cutoff \cite{Giamarchi}.
In the zero-temperature case considered here the evaluation of the
correlation function ${\cal G}(X)$ amounts to normal ordering the
product $\psi_2(x_L,t)\psi^{\dagger}_2(x_R)$ taking into account
Eqs.\ (\ref{Solution}). We finally obtain the following result:
\begin{align}
\log {\cal G}= &- \sum_n\frac{\pi}{4\omega_n}
\left\{|F_n(x_L)|^2+|F_n(x_R)|^2\right.
\nonumber \\
&- \left. 2F_n^*(x_L)F_n(x_R)e^{-i\omega_nt}\right\},
\label{Correlator}
\end{align}
where $F_n(x)= \Phi_n(x)+i\Theta_n(x)$.

{\it Scattering of plasmons.}---We note that the result
(\ref{Correlator}) holds for arbitrary potentials $U$ and $V$.
However, in our model the interaction is localized  between points
$x_L$ and $x_R$, therefore the correlator (\ref{Correlator}) can be
expressed in terms of the scattering properties of plasmons. Indeed,
in an open system, the differential equations (\ref{Motion})
describe the scattering of incoming plane waves with continuous
spectrum $\omega=v_Fk>0$ to outgoing plane waves. The scattering
matrix is symmetric, therefore quite generally we can write for the
transmission coefficient ${\cal T}=|{\cal T}|e^{i\varphi}$, and for
the reflection coefficients ${\cal R}=i|{\cal
R}|e^{i(\varphi+\delta)}$ and ${\cal R}'=i|{\cal
R}|e^{i(\varphi-\delta)}$, where $\varphi$ and $\delta$ are
scattering phases. Imposing now the periodic boundary condition on
the interval $[0,W]$, we obtain a discrete set of eigenfunctions
$\{\Phi_n,\Theta_n\}$, $n=0,\pm1,\ldots$, which take the following
form outside the scattering region:
\begin{equation}
\Phi_n(x)=\sqrt{\frac{2v_F}{W}}\times
\begin{cases}
\sin\left[k_nx-\frac{1}{2}(k_nW+\delta)\right], & n>0,\\
\cos\left[k_nx-\frac{1}{2}(k_nW+\delta)\right], & n<0,
\end{cases}
\label{fpm}
\end{equation}
and $\Theta_{n}(x)=\Phi_{-n}(x)$. They are normalized according to
equation (\ref{Normalization}), and the spectrum is given by
$k_nW=\left|2\pi n+\arccos|{\cal T}|-\varphi\right|$.

Substituting now $\Phi_n$ and $\Theta_n$ from Eq.\ (\ref{fpm}) into Eq.\
(\ref{Correlator}) and taking the limit $W\to\infty$, we finally express
the correlation function of electrons in terms of the transmission coefficient ${\cal T}$
for plasmons:
\begin{equation}
\log {\cal G}(X)= - \int_0^\infty\frac{dk}{k}
\left[1-{\cal T}^*(k)e^{-iXk}\right],
\label{Result}
\end{equation}
where, we remind, $X\equiv x_R-x_L+v_Ft$. This equation together
with Eqs.\ (\ref{Visibility-phase}) is one of the central results of
our paper.
In the non-interacting case the transmission is
perfect, ${\cal T}=1$, and equation (\ref{Result}) generates the
correlator ${\cal G}=-iv_Fn_F/(X-i0)$ for free fermions. One obtains
$|{\cal G}_{\Delta\mu}|=2\pi n_F$, which implies [see Eqs.\
(\ref{Visibility-phase})]  that the transport is coherent for
arbitrary bias $\Delta\mu$. 
Conversely, in the case of
nonzero interaction ${\cal T}\to 0$ for large $k$, the correlator
${\cal G}$ becomes independent of $t$, and the visibility $V_{AB}$
vanishes for large bias $\Delta\mu$. 
Next, we consider
a simple and natural model of a long-range interaction,
which qualitatively reproduces the puzzling results of the experiment
\cite{Heiblum1}.

{\it Long-range interaction model.}---We assume capacitive coupling between
edge states: $H_{\rm int}=Q^2/2C$,
where $Q=\int_{0}^{L}dx\,(\rho_2+\rho_3)$ is the total
charge in the interaction region. Then $U=2/C$
inside the interval $[0,L]$, while $V=0$.
The equations (\ref{Motion}) are straightforward to solve, and
we obtain the transmission coefficient:
${\cal T}=1+(1/2)|D|^2/(D+i\pi \omega C)$, where $D=e^{ikL}-1$.

We are interested in the first few resonances, therefore $kL\sim 1$.
Then the second term in the denominator of ${\cal T}$ is of the
order of $v_F/e^2$, i.e.\ the inverse interaction constant. In the
quantum Hall system of \cite{Heiblum1} this value is much smaller
than 1, so the second term in the denominator can be neglected, and
we arrive at
\begin{equation}
{\cal T}=(1+e^{-ikL})/2.
\label{T1}
\end{equation}
It is quite remarkable that the interaction constant drops from
the final  result leading to the universality which will be
addressed below.

The evaluation of the integral
(\ref{Result}) is now straightforward, and we obtain the electron correlator
\begin{equation}
{\cal G}(X)=-iv_Fn_F\,[(X-i0)(X-L-i0)]^{-1/2}
\label{C}
\end{equation}
Evaluating further the Fourier transform of ${\cal G}$, we find that
${\cal G}_{\Delta\mu}=
2\pi n_F\exp(i\Delta\mu L/2v_F)J_0(\Delta\mu L/2v_F)$, where $J_0$ is the zero-order
Bessel function.
Finally, using Eqs.\ ({\ref{Visibility-phase}}) we obtain that (see Fig.\ \ref{Bessel}):
\begin{equation}
V_{\rm AB}/V^{(0)}_{\rm AB}=|J_0(\Delta\mu L/2v_F)|,
\label{Long-range}
\end{equation}
while $\varphi_{\rm AB}(\Delta\mu)$ exhibits $\pi$ jumps at the
zeros of the Bessel function, in full agreement with the experiment
\cite{Heiblum1}. The characteristic $L$-dependence of the position
of zeros of the visibility suggests that our theory may be
experimentally verified by changing the length $L$ of the resonator.

The key feature of the
model is the presence of strong jumps in the potential $U$ at the
points $x=0$ and $x=L$. The system compensates this effect by
adjusting $\Phi(0)=\Phi(L)$, which immediately gives ${\cal T}
e^{ikL}=1+{\cal R}$, and ${\cal T} =e^{-ikL}+{\cal R}'e^{ikL}$,
independent of details of the interaction. Solving these equations
we obtain ${\cal T}=\cos\gamma\, e^{i(\gamma-kL)}$, where the phase 
$\gamma$ depends on the interaction. In our capacitive model $U$ is constant, and
$2\gamma$ is equal to the phase $kL$, accumulated between $x=0$ and
$x=L$, which leads to the result (\ref{T1}). The phase $\gamma$ will
increase, if the realistic repulsive interaction is taken into
account. However, this should merely shift zeros of the visibility
downward, as compared to zeros of the Bessel function.

\begin{figure}
\epsfxsize=8cm
\epsfbox{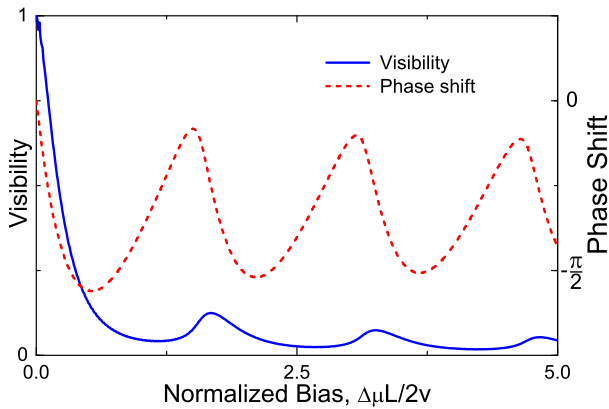}
\caption{The visibility for the case of a short-range interaction inside
the resonator (Luttinger constant $K=0.1$) is plotted versus bias normalized 
to the interaction dependent group velocity $v$ of plasmons. }
\vspace{-5mm}
\label{Luttinger-plot}
\end{figure}

{\it Short-range interaction model.}---To complete our analysis, we
investigate the case of short-range interactions at the edge, and
therefore write $U=2\pi U_0\delta(x-y)$, and $V=2\pi
V_0\delta(x-y)$. Then the Hamiltonian acquires the standard form
\cite{Giamarchi} $H_{\rm LL}=(1/2\pi)\int
dx[(v/K)(\nabla\phi)^2+vK(\nabla\theta)^2]$. The jump in the group
velocity $v$ and the Luttinger constant $K$
at the points $x=0$ and $x=L$ leads to resonant scattering of plasmons.
The result for the electronic correlator may be
presented as follows
\begin{equation}
{\cal G}=-iv_Fn_F\,\prod_{n=0}^\infty[X+X_n-i0]^{-\alpha_n},
\label{C2}
\end{equation}
where $X_n=[(2n+1)(v_F/v)-1]L$, and $\alpha_n=4K/(K+1)^2\times
[(K-1)/(K+1)]^{2n}$. The infinite product in (\ref{C2}) is due to
multiple scattering at the ends of the resonator and $X_n$ are the
lengths of corresponding paths. The result of the numerical
evaluation of the Fourier transform for ${\cal G}_{\Delta\mu}$ is
shown on Fig.\ \ref{Luttinger-plot}. Note that the lobe-type
structure in the visibility is absent and the phase shift develops
smooth oscillations. This behavior disagrees with experimental
findings.

To summarize, we have demonstrated that the lobe structure and phase
slips observed in Ref.\ \cite{Heiblum1} provide evidence of strong
long-range interaction between quantum Hall edge states. By
comparing two different models of edge states, we have demonstrated
strong sensitivity of AB oscillations to the character of the
interaction. This suggests that AB interferometry can be used as a
powerful tool for the tomography of interactions at the quantum Hall
edge and possibly in other 
systems. 

We acknowledge support from the Swiss NSF. V.C.\ is grateful for the
hospitality of workshop COQUSY06 at the MPI, Dresden.

\bibliographystyle{apsrev}

\begin{thebibliography}{99}
\bibitem{footnote}
In Fig.\ 1 of  Ref.\ \cite{Heiblum1} the counter-propagating edge
state goes from the source S3 to QPC0 and under the air bridge comes
very closely to the upper arm of the interferometer, which is the
part of the channel between QPC1 and QPC2.
\bibitem{footnote1}
We believe that the physics we discuss is not
determined by generally complex
microscopic properties of QPCs.
\bibitem{footnote2}
Since in our model the interaction is localized inside the interferometer,
the perturbation theory in tunneling is free from infrared divergences 
\cite{Giamarchi}.
\end{thebibliography}

\end{document}